# Multimodal pathophysiological dataset of gradual cerebral ischemia in a cohort of juvenile pigs

Martin G. Frasch[1], Bernd Walter[2,3], Christophe L. Herry[4], and Reinhard Bauer[3]

[1] University of Washington School of Medicine, Center on Human Development and Disability, Seattle, WA, USA
[2] Department of Spine Surgery and Neurotraumatology, SRH Waldklinikum, Gera, Germany
[3] Institute of Molecular Cell Biology, Jena University Hospital, Jena, Germany
[4] Dynamical Analysis Lab, Ottawa Hospital Research Institute, Ottawa, ON, Canada

**Corresponding author:**

Dr. Martin G. Frasch

Department of Obstetrics and Gynecology

University of Washington

1959 NE Pacific St

Box 356460

Seattle, WA 98195

Email: mfrasch@uw.edu

## Abstract

Ischemic brain injuries are frequent and difficult to detect reliably or early. We present the multi-modal data set containing cardiovascular (blood pressure, blood flow, electrocardiogram) and brain electrical activities to derive electroencephalogram (EEG) biomarkers of corticothalamic communication under normal, sedation and hypoxic/ischemic conditions with ensuing recovery. We provide technical validation using EEGLAB. We also delineate the corresponding changes in the electrocardiogram (ECG)-derived heart rate variability (HRV) with the potential for future in-depth analyses of joint EEG-ECG dynamics. We review an open-source methodology to derive signatures of coupling between the ECoG and electrothalamogram (EThG) signals contained in the presented data set to better characterize the dynamics of thalamocortical communication during these clinically relevant states. The data set is presented in full band sampled at 2000 Hz, so the additional potential exists for insights from the full-band EEG and high-frequency oscillations under the bespoke experimental conditions. Future studies on the dataset may contribute to the development of new brain monitoring technologies, which will facilitate the prevention of neurological injuries.

## Background & Summary

Surface EEG contains information about corticothalamic communication, which can be quantified even without invasive insertion of thalamic electrodes.[1,2] We present a unique data set from our laboratory[1,2] along with an approach to derive EEG biomarkers of corticothalamic communication under normal, sedation and hypoxic/ischemic conditions.

We hypothesize the proposed biomarkers of corticothalamic communication derived from less than 10 minutes of EEG data will be sensitive to early abnormal ECoG states (i.e., surface ECoG will predict changes in EThG).

We hope the present data set will lay the foundation for new brain monitoring technologies, which will facilitate the prevention of neurological disorders. The data set is derived from a series of complex physiological experiments in juvenile pigs.

First, we present a method to acquire and the data set containing the electrothalamographic (EThG) and electrocorticographic (ECoG) data in juvenile pigs undergoing various sedation regimes followed by gradual ischemia and recovery periods. Accompanying cardiovascular time series including blood pressure and electrocardiogram recordings are also provided. Second, we present an elementary approach to quantify the ECoG activity on the cohort level identifying state-specific independent component (IC) features of EEG common to all animals studied.

All experiments were carried out in accordance with the European Communities Council Directive 86/609/EEC for animal care and use. The Animal Research Committee of the Thuringian State government approved laboratory animal protocols.

**The choice of this animal model** is dictated by its amenability to complex stereotactic chronic instrumentation, monitoring studies of sedation and clinically relevant patterns of hypoxic/ischemic injury in a relatively large and gyrencephalic brain.[2]

## Methods

### 1) Instrumentation

**General instrumentation.** 13 female juvenile pigs (7 weeks of age) were instrumented as published.[1,2] Ventilation was performed in a pressure-controlled mode and was controlled by end-expiratory $CO_2$ monitoring and hourly arterial blood gas analysis. Body temperature was maintained at 37.5±0.5°C. The urinary bladder was punctured and drained. Arterial blood pressure (ABP) was monitored continuously using catheters that were introduced into the abdominal aorta via the saphenous arteries. A left thoracotomy was then performed through the third intercostal space. The pericardium was carefully opened, and a cerclage of plastic-coated wire was performed around the trunk of the pulmonary artery to appropriately adjust the trunk diameter for cardiac output control during the induction and maintenance of gradual ischemia. An electrocardiogram (ECG) was recorded from standard limb leads using stainless steel needle electrodes.

**Instrumentation of the head (Fig. 1).** After fixing the head in a stereotactic frame, the skull was exposed and trepanations were made for the insertion of ten electrocorticographic electrodes for bilateral leads from frontal, parietal, central, temporal, and occipital regions and guides for electrodes into the left thalamus, specifically, the reticular (RTN) and dorsolateral (LD) nuclei. All trepanations were then sealed with bone wax and dental acrylic. RTN was chosen as a non-specific nucleus with a high density of GABA A receptors responsible for the sedative effects of propofol [3,4] and gating influence on thalamocortical oscillations [5,6], while LD represented a specific exteroceptive nucleus. Between RTN, LD and eight ECoG channels we captured representative specific and non-specific thalamocortical communication.

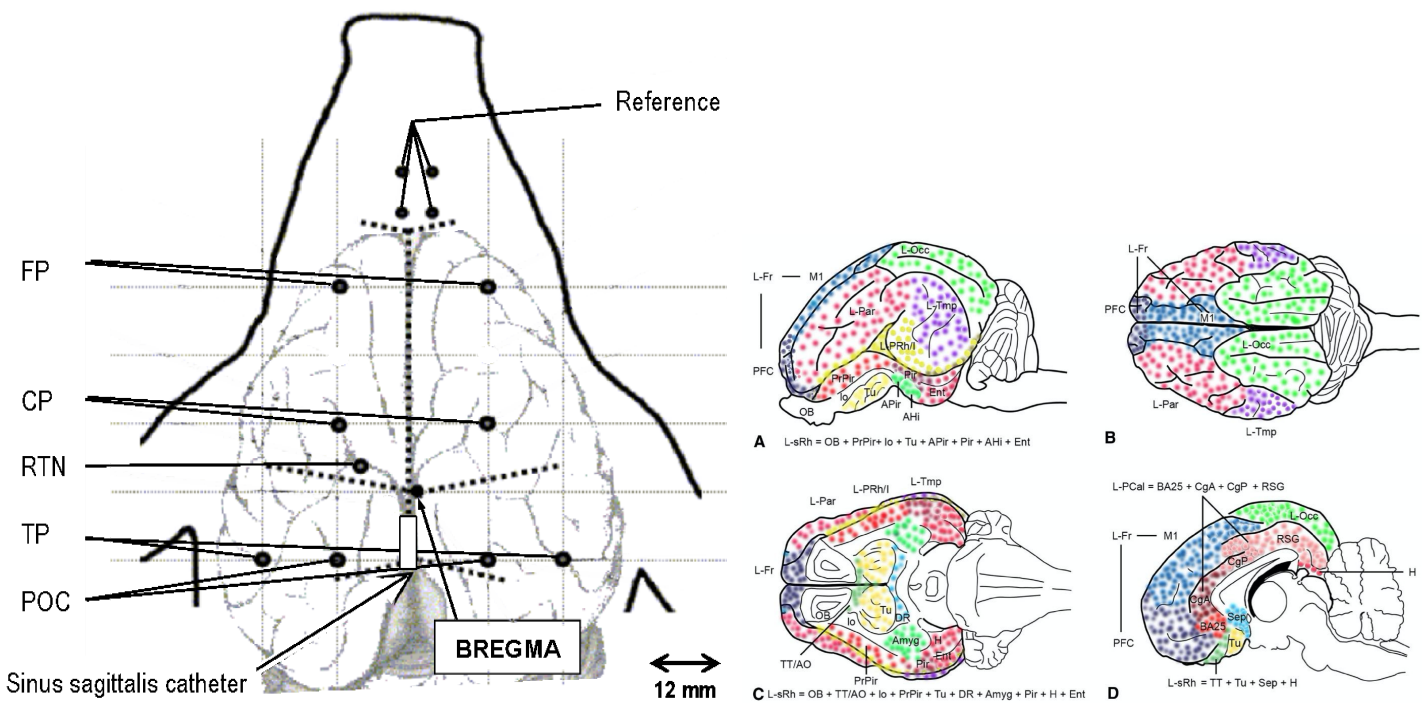

**Figure 1. TOP LEFT: Instrumentation of pig for recording ECoG and EThG.** For details see Table 1. **TOP RIGHT**: Schematic drawings of the Goettingen minipig telencephalon's lobes and some of the major subareas online on https://www.cense.dk/ [7] **A. Lateral view**: visible frontal lobe subareas (L-Fr) are indicated separately. **B. Dorsal view**: two frontal lobe subareas are indicated separately. **C. Ventral view**: visible subareas of the subrhinal lobe are visualized schematically. **D. Midsaggital view**: the visible subareas of the frontal, pericallosal and subrhinal lobes are visualized separately.

AHi, amygdalohippocampal transition area, Amyg amygdala, AO anterior olfactory nucleus, APir amygdalopiriform transition area, BA25 subgenual cortex, CgA anterior cingulate cortex, CgP posterior cingulate cortex, DR diagonal region, Ent entorhinal cortex, H hippocampus, L-Fr frontal lobe, lo lateral olfactory tract, L-Occ occipital lobe, L-Par parietal lobe, L-PRh/I perirhinal/insular lobe, L-sRh subrhinal lobe, L-Tmp temporal lobe, M1 motor cortex, OB olfactory bulb, PFC prefrontal cortex, Pir piriform cortex, PrPir prepiriform cortex, RSG retrosplenial granular cortex, Sep septum, Tu olfactory tubercle, TT taenia tecta.

**Table 1. Stereotactic coordinates*.**

| Position | Lateral of sagittal suture | Anterior (a) / Posterior (p) of bregma | Depth from the dura | Equivalent according to 10/20 single plane projection of the head [8] |
|---|---|---|---|---|
| Frontal ECoG, Ch. 1 & 2 (left & right) | 12 mm | a 30 mm | | Fp1, Fp2 |
| Parietal ECoG, Ch. 3 & 4 (left & right) | 12 mm | a 20 mm | | F3, F4 |
| Central ECoG, Ch. 5 & 6 (left & right) | 12 mm | a 10 mm | | C3, C4 |
| Temporal ECoG, Ch. 7 & 8 (left & right) | 24 mm | p 10 mm | | T5=P7, T6=P8 |
| Occipital ECoG, Ch. 9 & 10 (left & right) | 12 mm | p 10 mm | | P3, P4 |
| EThG, Ch. 11 (RTN = Nucl. reticularis thalami) | 9 mm | a 2 mm | 24 mm | |
| EThG, Ch. 12 (LD = Nucl. dorsolateralis thalami) | 5 mm | p 2 mm | 20 mm | |

* Reference: Nz

## 2) Description of experimental stages

After post-surgical recovery, the pigs were allowed 90 minutes to stabilize (**State 1**) with electrocorticogram (ECoG) and electrothalamogram (EThG) recorded continuously until necropsy. Then, isoflurane in $N_2O$ and $O_2$ was discontinued and ventilation with 100% $O_2$ was performed for 5 minutes. Another phase (**State 2a**) ensued in which intravenous bolus injection of fentanyl (0,015 mg/kg b.w.) was carried out followed by continuous iv infusion of fentanyl (0.015 mg/kg b.w./h) for 90 minutes (**State 2b**). Next, individual doses of propofol required for the maintenance of deep anesthesia were determined under continuous control of mean ABP (MABP). Propofol was infused intravenously (0.9 mg/kg BW/min for ~ 7 min) until a burst suppression pattern (BSP) appeared in the ECoG. The depth of anesthesia was subsequently maintained for 25 minutes via propofol administration (~ 0.35 mg/kg b.w./min) (**State 3**). Next, 30% of the propofol dose required for BSP induction was continuously administered over the course of 90 minutes to produce moderate anesthesia (**State 4**). **States 5** represented the measurement 60 min after induction of the moderate propofol sedation.

We induced gradual cerebral ischemia as follows.[9] First, the cisterna magna was punctured by a lumbar puncture needle that was fixed in place by dental acrylic resin for elective artificial cerebrospinal fluid infusion/withdrawal to control ICP. Then, the mean ABP was adjusted to about 90 mmHg by the appropriate curbing of the pulmonary trunk diameter with the plastic-coated cerclage. The cerebral perfusion pressure (CPP) was then decreased at 25 mmHg, which was calculated as the difference between MABP and the intracranial pressure (ICP) by appropriated elevation of the ICP. The increase in the ICP was achieved by the infusion of artificial cerebrospinal fluid (warmed to 37°C) into the subarachnoid space via the punctured cisterna magna. The Cushing response during severe brain ischemia was prevented by the appropriate curbing of the pulmonary trunk diameter with the plastic-coated cerclage to control cardiac output. Finally, the cerclage was opened completely, and the artificial cerebrospinal fluid was withdrawn to reach an ICP < 10 mmHg.

The state of gradual ischemia was maintained for 15 min (**State 6**). This was followed by 60 min of recovery (**States 9-12**).

At the end of the experiment, the animals' brains were perfusion-fixed.[2] Afterward, the head was removed, immersion-fixed, the brain was removed and electrode positions were visually and histologically confirmed.

**Data acquisition and analyses.** Unipolar ECoG and EThG were amplified, filtered, fed into a PC using a 25 channel A/D board at 2000 Hz and stored.

**Cardiovascular and metabolic parameters.** ECG, heart rate, MABP, body temperature, arterial and brain venous pH, $pCO_2$, $pO_2$, $O_2$ saturation, glucose, lactate, and hemoglobin values were measured at each State as published.[1,2]

**Cerebrovascular and cerebral metabolic parameters.** At states 1,2,5,6 and 12, whole and regional brain blood flows were measured with colored microspheres[10] together with brain oxygen extraction (arterial – sagittal sinus blood oxygen content difference, $AVDO_2$) and cerebral metabolic rate of oxygen (the product from cortical blood flows and $AVDO_2$).[9]

## 3) Signal analysis methodology

EEGLAB v2019.1 for Linux was used in Matlab R2013b (MathWorks, Natick, MA) to analyze ECoG/EThG data.[11] The continuous individualized multi-organ variability analysis (CIMVA) was used to analyze the ECG-derived heart rate variability (HRV) as reported; CIMVA renders a CIMVA HRV measures matrix.[12,13]

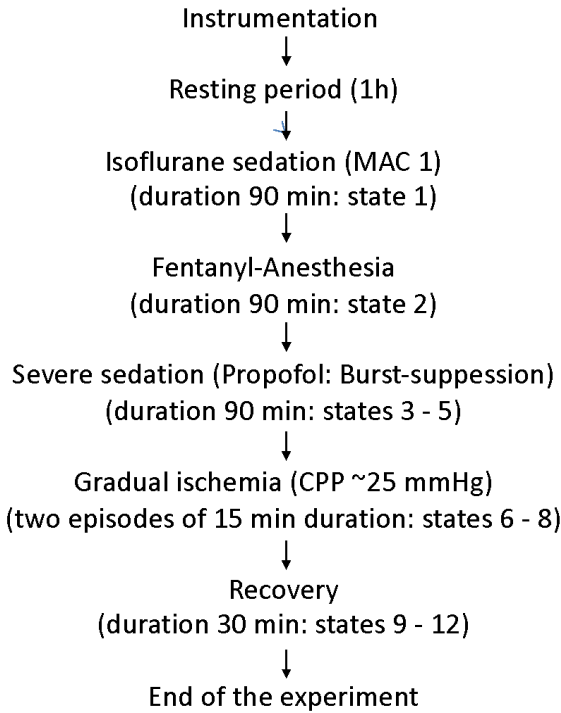

**Figure 2. Experimental protocol of pig model of sedation and gradual ischemia.**

## Data records

The data structure and annotation are as follows. There are 25 channels at a sampling rate of 2000 Hz and recording duration of 300 s per state each containing the following channels.

1. ABP, arterial blood pressure measured from brachial artery
2. ABP_Tip, arterial blood pressure measured in external carotid artery [higher quality signal than Ch. 1 due to absent signal dampening]
3. ECG, electrocardiogram
4. CO2 con, the $CO_2$ concentration of the ventilatory gas
5. Airpres, airway pressure
6. ICP, intracranial pressure
7. CVP, central venous pressure
8. EMF (blood flow: left carotid)
9. EMF2e (blood flow: right carotid)
10. Tc, brain tissue $pO_2$
11. Tc temp, brain tissue temperature in Celsius
12. **Channels 12 - 22**: ECoG 1 – 10; Channel 11: EThG RTN, Channel 12: EThG LD nucleus
13. Trig SEP or Trig AEP [empty in spontaneous recordings]
14. Temp re, rectal temperature in Celsius

The sample raw recording is shown in Fig. 3.

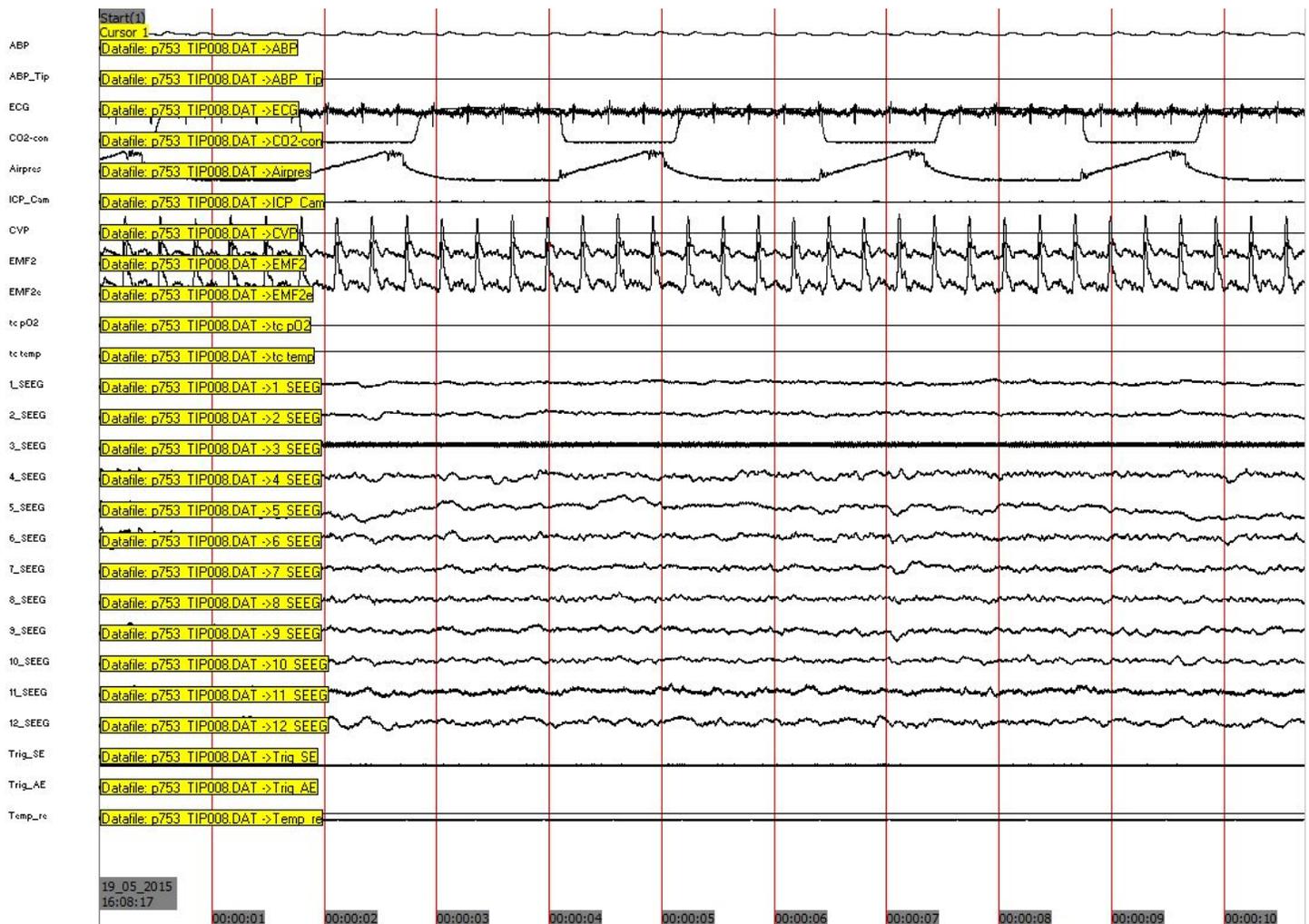

**Figure 3. Representative raw recording during data acquisition in Watisa software.**

All animals underwent gradual ischemia. The two groups are defined with respect to their exposure to propofol sedation. The experimental group is comprised of N=6 animals (P728, 737, 738, 743, 746, 752). They experienced propofol burst suppression followed by moderate propofol sedation prior to gradual ischemia. The propofol sham group is comprised of N=5 animals (P739, 749, 753, 791, 794). They did not receive propofol. Instead, they were continued on fentanyl analgosedation prior to gradual ischemia.

The animal P728 experienced an experimental mishap during the first gradual ischemia stage (state 6). The recordings are presented up until state 4 and can be grouped with other animals' data for the respective states 1 through 4. The animal P738 demised prematurely during the second gradual ischemia (state 8). Consequently, the data is presented up until state 6 and can also be grouped and studied together up until this point. Here, special consideration should be made for the potential incipient deterioration leading up to the early demise which represents a point of interest.

File duration per state is 300 s. Sampling rate: 2000 Hz. Note that ECoG channel 3 had noise in some instances, so it needs to be discarded from analyses where appropriate. We observed that the signal in channel 3 can be recovered with independent component analysis in some instances.

**Table 2. Review of experimental stages and the respective available data sets***

| Experimental stage number | Experimental stage | |
|---|---|---|
| **state1** | Isoflurane | P_728, **P_737,** P_738, **P_739**, P_743, P_746, P_749, **P_752, P_753**, **P_791, P_794** |
| *state2* | Fentanyl | P_728, **P_737,** P_738**, P_739**, P_743, P_746, P_749, **P_752,** *P_753*, **P_791, P_794** |
| state2.5 | 90 min post-fentanyl | P_739, P_753 |
| **state3** | Propofol | P_728, **P_737,** P_738**,** P_743, P_746, **P_752** |
| **state4** | Moderate sedation - immediate measurement | P_728, **P_737,** P_738**,** P_743, P_746, **P_752** |
| **state5** | 60 min post-moderate sedation | **P_737**, P_743, P_746, **P_752** |
| state5.5 | Fentanyl directly pre-ischemia | P_794 |
| *state6* | 1st ischemic phase: gradual ischemia | P_738, **P_739**, P_746, P_749, **P_752,** *P_753* |
| *state8* | 2nd ischemic phase: gradual ischemia | **P_739, P_752, P_753, P_791, P_794** |
| state9** | 15 min recovery post-ischemia | **P_737**, P_743 |
| state9a** | 15 min recovery post-ischemia | P_746 |
| **state9b*** | 15 min recovery post-ischemia | **P_739**, P_746, P_749, **P_752, P_753, P_791, P_794** |
| **state10** | 30 min recovery post-ischemia | P_746, P_749, **P_752, P_753, P_791, P_794** |
| **state11** | 45 min recovery post-ischemia | **P_737**, **P_739**, P_743, P_749, **P_752, P_753, P_791, P_794** |
| *state12* | 60 min recovery post-ischemia | **P_737,** P_743, P_749, **P_752,** *P_753,* **P_791, P_794** |

\* Bold font: animals and states selected for the analyses presented in the Technical validation section.
\*\* When there are two "ischemic" phases and two corresponding "15 min recovery post-ischemia" phases, the first recovery phase is denoted as "state9a" and the second, "state9b".

In addition to the raw data, we deposited the corresponding, time-matched cerebral-venous measurements of blood gases, electrolytes, metabolites, cardiovascular and cerebrovascular as well as CMRO2 data. The exact list of the types of data is outlined in Table 3. All data have been deposited on Figshare: 10.6084/m9.figshare.7834442.

**Table 3. Parameters collected during each state of the experiment.**

| | |
|---|---|
| Arterial blood data | body temperature [°C] |
| | hemoglobin [mmol/l] |
| | oxygen saturation[%] |
| | pH |
| | $pCO_2$ [mmHg] |
| | $pO_2$ [mmHg] |
| | acid base excess [mmol/l] |
| | $K^+$ |
| | $Na^+$ |
| | glucose |
| | lactate |
| | Carboxy-hemoglobin (%) |
| | Methemoglobin (%) |
| | Oxygen content (Vol%) |
| Sagittal sinus venous blood data | body temperature [°C] |
| | hemoglobin [mmol/l] |
| | oxygen saturation[%] |
| | pH |
| | pCO2 [mmHg] |
| | pO2 [mmHg] |
| | acid-base excess [mmol/l] |
| | K$^+$ |
| | Na$^+$ |
| | glucose |
| | lactate |
| | Carboxy-hemoglobin (%) |
| | Methemoglobin (%) |
| | Oxygen content (Vol%) |
| Cardiovascular data | arterial blood pressure (mm Hg) |
| | cardiac output ((ml/min * kg body weight) |
| Cerebrovascular data | CBF_Whole brain (ml/100g*min) |
| | cerebral perfusion pressure (mm Hg) |
| | cerebral metabolic rate of oxygen (ml/100g*min) |
| Brain regional blood flow (ml/100g*min) | Thalamus |
| | Frontal cortex |
| | Parietal cortex |
| | Temporal cortex |
| | Occipital cortex |

## Technical validation

### 1) Brain electrical activity

We present the approach and findings of signatures of global and brain-regional changes in ECoG-derived independent component activity during sedation, ischemia and recovery states. Table 2, bold font, lists the animal IDs and experimental states we selected for technical validation. We considered states 1-5 combined as sedation, states 6 and 8 as ischemia and states 9b - 12 as post-ischemic recovery. Using STUDY design feature of EEGLAB in Matlab, we conducted statistical analyses on a group of six animals (P737, P739, P752, P753, P791, P794) as follows:

1) read raw data (2,000 Hz sampling rate);
2) select 10 ECoG channels;
3) remove DC offset;
4) resample at 100 Hz;
5) a bandpass filter with FIR 1 - 40 Hz and save as *.set files (also shared on figshare)
6) map ECoG channel locations for better visualization using Nz as reference (.loc file shared on FigShare; cf. Table 1)
7) create STUDY in EEGLAB; this approach permits statistical level inferences across animals for state-specific changes common to all subjects
8) compute independent components for all animals and states in the STUDY
9) identify IC clusters (using k-means approach) that differentiate sedation, ischemia, and recovery states.

We deployed EEGLAB software suite (http://sccn.ucsd.edu/eeglab)[11,14,15] available as Matlab / Octave add-on or a pre-compiled open-source package for Windows, Mac and Linux operating systems to conduct technical validation of our data set and to demonstrate some initial findings of interest for future studies. The technical advantage of this approach is that this software package is open source and readily available online for the major operating systems.

We focused on the representative experimental states 2 (baseline), 6 (acute gradual ischemia) and 12 (60 min recovery) for all subjects using the STUDY functionality of EEGLAB which allows studying all subjects at once compared the event-related potential (ERP) responses within the group. ERP represented the response to the states of the experiment (Table 2). To facilitate the computation, we downsampled the data to 100 Hz and focused on the ECoG channels for EEGLAB-based analysis which allowed us to map the channels according to 10/20 using a channel location file (see Figshare). Future studies on this data set could consider the full bandwidth ECoG/EThG information contained in the recordings and include EThG channels in the investigations.

We performed the independent component analysis (ICA) on animals identified in Table 2 followed by wavelet time-frequency and cross-coherence analyses on the identified independent components.

The results are presented in Figure 4. The analysis of the raw data revealed the activity to be concentrated in the delta - alpha band range which experienced most of the reduction in spectral power in the individual channels as well as in cross-coherence due to ischemia and during recovery. Following ICA, we observed an overall reduced cross-coherence in IC2 - IC1 compared to raw channel C3 - C4 analysis, but still following a similar pattern throughout states 5, 6 and 12 (Fig. 4, panels G and H). As expected, a portion of the observed coherence is accounted for by volume conduction effects which are removed by ICA. As evidenced by the group analysis results of which are shown in Fig. 4, panel I, 60 min recovery did not suffice to restore the global reductions spectral power activity seen during the ischemic states.

Ischemia reduces total power across the spectrum going from sedation states over to ischemia and then recovery, however, during recovery the bihemispheric coupling is re-established. A combination of spectral power characteristics and coupling dynamics can be used to distinguish ischemia-induced loss in spectral power and post-ischemic recovery. Clustered ICA also distinguishes these states.

A

B
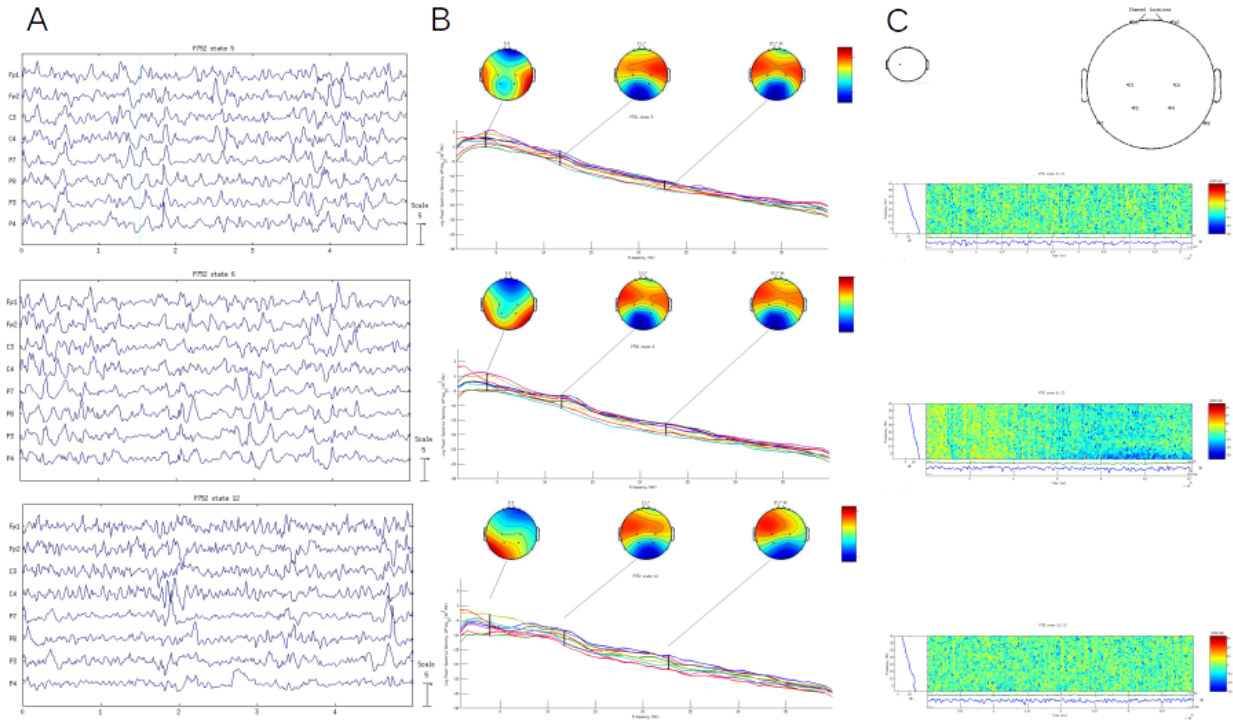

C

D
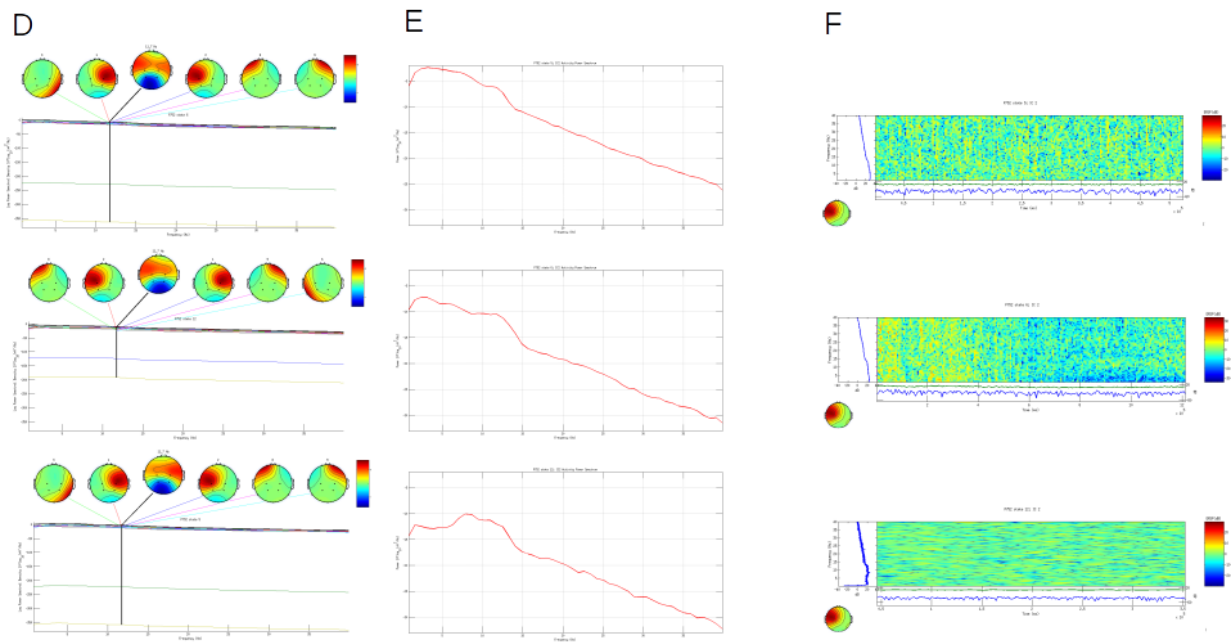

E

F

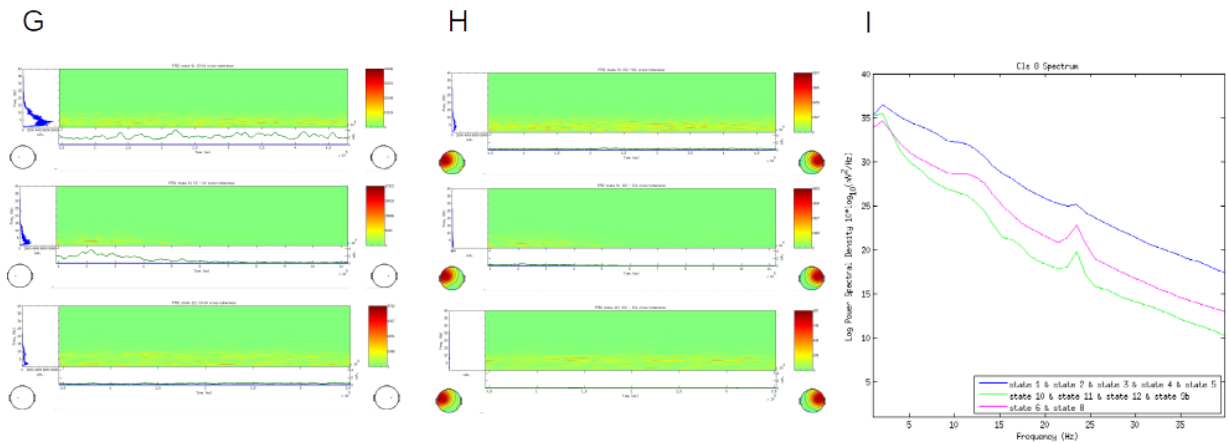

**Figure 4. Representative findings (A - H) and group-level overview (I) showing the separation of three states: sedation, ischemia and recovery**, in the set of five animals (P737, P752, P753, P791, P794) based on nine clustered independent components computed from eight ECoG channels (cf. topological map in panel C, top right).*

In the example of P752, we focus on the representative states 5 (60 min into moderate propofol *sedation*), 6 (gradual *ischemia*) and 12 (60 min post-ischemic *recovery*). For each state, we show the following:

1) the raw ECoG tracings (panel A, normalized and DC removed), corresponding topological power spectrum maps (panel B) showing gradual suppression of activity with peaks at 4, 12 and 23 Hz and the time-frequency representation using wavelet transform on the channel 5 (i.e., C3; panel C);
2) the independent component (IC) analyses (panel D) with focus on 12 Hz activity (one of the clearest consistent peaks in the entire group) for the five top ICs in each state; representative IC2 power spectrum (panel E), again showing the overall reduction of activity and the corresponding time-frequency representation using wavelet transform on the IC2 (panel F);
3) cross-coherence representations of bihemispheric coupling between C3-C4 (panel G), IC2 - IC1 (panel H)
4) Group analysis (panel I) showing the gradual reduction of spectral power between the experimental states representing sedation (states 1 - 5, blue line), ischemia (states 6 and 8, red line) and recovery (states 9b - 12, green line). We observe peaks around 12 and 23 Hz which triggered our focus on these frequencies in the representative examples shown in panels A - H.

*Note that all axes are set identically for easy comparison except the topological and time-frequency transform heat maps. These are individually optimized in range for best viewing experience.

Overall, this finding shows that ischemia and recovery states can be distinguished globally using such an ICA approach. As a proof-of-concept, this analytical approach in EEGLAB demonstrates the potential of the presented experimental data set to yield new pathophysiological insights into global and brain-regional responses to sedation, gradual ischemia, and post-ischemic recovery.

### 2) Cardiovascular activity

To demonstrate an approach to use the present dataset for studying the unique features of cardiac electrical activities during sedation, ischemia and recovery stages, we exported the ECG channel (Ch. 3 or Ch.4) at 1000 Hz as EDF files for the same animals for which we presented the above approach on the ECoG data (see Figshare).

First, we computed multi-dimensional HRV measures from the ECG channel using the established CIMVA approach which renders a CIMVA HRV measures matrix.[12,13] Next, we grouped the findings according to the three main states of the experiments into sedation, ischemia and recovery following the same approach as in Fig. 4I. Finally, we visualized the changes in the CIMVA HRV matrix in Fig. 5 using the principal component analysis (PCA). The interactive version of Fig. 5 can be viewed here. The contribution of each HRV measure to the principal components is shown here interactively. PCA demonstrates how the HRV matrix dynamics separate the states of sedation, ischemia, and recovery.

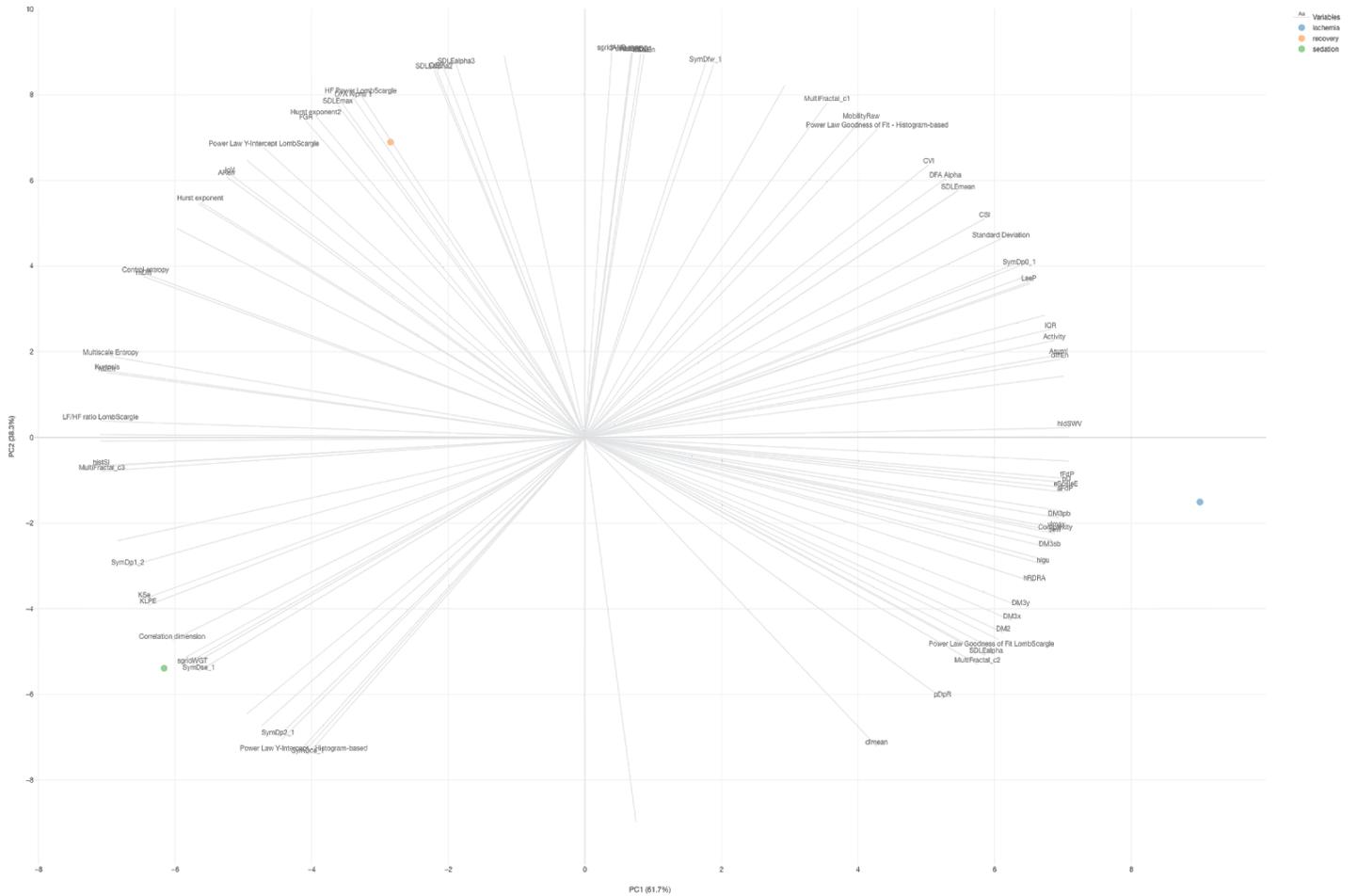

**Figure 5. Group level overview of the behavior of the ECG-derived multi-dimensional heart rate variability (HRV) measures**. Principal component (PC) analysis can be used to separate the states of sedation, gradual ischemia and recovery: green dot indicates sedation-related HRV measures; blue dot indicates the ischemia-related HRV measures; orange dot indicates the recovery-related HRV measures. The interactive version of Figure 5 can be viewed here. PC1 and PC2 explained 100% of the variability in the data.

## Usage notes

For thalamocortical and cortico-cortical communication, sedation-dependent linear and nonlinear coupling dynamics have been reported[1], but not yet characterized under conditions of gradual ischemia as a potential biomarker of brain state and recovery. Studies in rats have shown that global brain hypoxia-ischemia affects long-term information processing in thalamic circuitry and the transfer of sensory information in thalamocortical networks.[16,17] Newborn mice exposed to ischemic insult also suffer from the increased vulnerability of thalamocortical circuitry.[18] There is less data on thalamocortical responses to ischemia from larger mammals with stronger resemblance of brain maturity, developmental profile and injury patterns, such as sheep or pig.[19] The present dataset may help to close this data gap and yield new insights into monitoring, early detection, recovery of ischemic and post-ischemic brain states, in particular, thalamocortical communication which are important to help restore long-term brain health.

Anesthesia-induced changes in brain electrical activity, in particular, due to the dedicated GABA A receptor-mediated effects of propofol on the RTN, have been used to model and study changes in consciousness and behavioral state activity.[20–23] In this experimental design, we benefit from propofol sedation with electrode placement in the RTN, which, as discussed, is particularly amenable to linking drug-induced agonism on GABA A receptors of RTN and the sedation depth.[1] The combination of propofol sedation with subsequent ischemia data from ECoG and EThG, including Nucl. reticularis thalami, yields a rich dataset to study patterns of thalamocortical communication under conditions of sedation, ischemia and recovery.

The present dataset has been acquired at a 2,000 Hz sampling rate and is hence amenable to studies of the properties of high-frequency oscillations under conditions of various sedation regimes, gradual ischemia and recovery periods.[24–26] Of great interest, thanks to EThG recordings, it may be possible to relate the ECoG patterns of spontaneous high-frequency oscillations to their thalamic contributions in EThG.

In this data-oriented manuscript, we used EEGLAB for demonstration of the technical rigor, data quality and reproducibility. While rich in functionality and open source, in future studies of this dataset the EEGLAB application could be well-complemented by additional analytical approaches, also available open-source, such as the JIDT software package.[27] (https://github.com/jlizier/jidt/) JIDT is Java-based, offers GUI, requires no installation and runs on all major platforms. It also can be integrated with Matlab, Python or R, among others. Features, of interest to this dataset that are available with JIDT, include mutual information and transfer entropy. It would be of interest to compute these before and after ICA as well as after subtracting IC expecting the bi-channel / bivariate information measures to decrease after ICA and re-increase after removing ICs. Using JIDT software, one can also determine cross-entropy measures for ECoG - EThG channels prior to and after IC computation.

Finally, the simultaneous availability of the cardiovascular and brain electrical data lends itself to studying joint dynamics of, for example, ECG and ECoG or EThG. The relationships between these time series have been reported[28,29], sporadically, but much remains to be done to establish the physiological framework for these relations and to explore the biomarker potential of such multivariate EEG - ECG analyses. We leave this exciting direction of research to future studies using this data set.

## Code availability
EEGLAB has been used which is available as open-source. CIMVA documentation is available online. No proprietary code has been deployed in this study.

## Acknowledgments
The authors thank Konstanze Ernst, Rose Zimmer und Lothar Wunder for their skillful assistance during the experimentation and data analyses. We also gratefully acknowledge the many fruitful conversations with the late Professor Ulrich Zwiener, of good memory, who was an inspiration to us all on this and many other projects.

## Author contributions
R.B. and B.W. conceived, designed, and conducted the experiments and carried out the data analysis. M.G.F. designed and conducted the experiments. M.G.F. and C.L.H. carried out the analysis. All authors contributed to the interpretation of the data and drafting of the manuscript. All authors contributed to critical revision and approved the final version of the manuscript.

## Competing interests
The authors have no conflicts of interest to disclose.

## References

1.  Frasch, M. G. et al. Detecting the signature of reticulothalamocortical communication in cerebrocortical electrical activity. *Clin. Neurophysiol.* **118**, 1969–1979 (2007).
2.  Frasch, M. G. et al. Stereotactic approach and electrophysiological characterization of thalamic reticular and dorsolateral nuclei of the juvenile pig. *Acta Neurobiol. Exp.* **66**, 43–54 (2006).
3.  Alkire, M. T. & Haier, R. J. Correlating in vivo anaesthetic effects with ex vivo receptor density data supports a GABAergic mechanism of action for propofol, but not for isoflurane. *Br. J. Anaesth.* **86**, 618–626 (2001).
4.  Sonner, J. M. et al. GABA(A) receptor blockade antagonizes the immobilizing action of propofol but not ketamine or isoflurane in a dose-related manner. *Anesth. Analg.* **96**, 706–12, table of contents (2003).
5.  Bazhenov, M., Timofeev, I., Steriade, M. & Sejnowski, T. J. Self-sustained rhythmic activity in the thalamic reticular nucleus mediated by depolarizing GABAA receptor potentials. *Nat. Neurosci.* **2**, 168–74. (1999).
6.  Steriade, M. The GABAergic reticular nucleus: a preferential target of corticothalamic projections. *Proc. Natl. Acad. Sci. U. S. A.* **98**, 3625–3627 (2001).
7.  Bjarkam, C. R., Glud, A. N., Orlowski, D., Sørensen, J. C. H. & Palomero-Gallagher, N. The telencephalon of the Göttingen minipig, cytoarchitecture and cortical surface anatomy. *Brain Struct. Funct.* **222**, 2093–2114 (2017).
8.  Klem, G. H., Jasper, H. H. & Elger, C. The ten±twenty electrode system of the International Federation.
9.  Walter, B. et al. Age-dependent effects of gradual decreases in cerebral perfusion pressure on the neurochemical response in swine. *Intensive Care Med.* **36**, 1067–1075 (2010).
10. Walter, B., Bauer, R., Gaser, E. & Zwiener, U. Validation of the multiple colored microsphere technique for regional blood flow measurements in newborn piglets. *Basic Res. Cardiol.* **92**, 191–200 (1997).
11. Delorme, A. & Makeig, S. EEGLAB: an open source toolbox for analysis of single-trial EEG dynamics including independent component analysis. *J. Neurosci. Methods* **134**, 9–21 (2004).
12. Seely, A. J., Green, G. C. & Bravi, A. Continuous Multiorgan Variability monitoring in critically ill patients--complexity science at the bedside. *Conf. Proc. IEEE Eng. Med. Biol. Soc.* **2011**, 5503–5506 (2011).
13. Herry, C. L. et al. Vagal contributions to fetal heart rate variability: an omics approach. *Physiol. Meas.* (2019) doi:10.1088/1361-6579/ab21ae.
14. Delorme, A., Palmer, J., Onton, J., Oostenveld, R. & Makeig, S. Independent EEG sources are dipolar. *PLoS One* **7**, e30135 (2012).
15. Delorme, A. et al. EEGLAB, SIFT, NFT, BCILAB, and ERICA: new tools for advanced EEG processing. *Comput. Intell. Neurosci.* **2011**, 130714 (2011).
16. Shoykhet, M. et al. Thalamocortical dysfunction and thalamic injury after asphyxial cardiac arrest in developing rats. *J. Neurosci.* **32**, 4972–4981 (2012).
17. Shoykhet, M. & Middleton, J. W. Cardiac Arrest-Induced Global Brain Hypoxia-Ischemia during Development Affects Spontaneous Activity Organization in Rat Sensory and Motor Thalamocortical Circuits during Adulthood. *Frontiers in Neural Circuits* vol. 10 (2016).
18. Liu, X.-B., Shen, Y., Pleasure, D. E. & Deng, W. The vulnerability of thalamocortical circuitry to hypoxic-ischemic injury in a mouse model of periventricular leukomalacia. *BMC Neurosci.* **17**, 2 (2016).
19. Morrison, J. L. et al. Improving pregnancy outcomes in humans through studies in sheep. *American Journal of Physiology-Regulatory, Integrative and Comparative Physiology* **315**, R1123–R1153 (2018).
20. Steriade, M. Synchronized activities of coupled oscillators in the cerebral cortex and thalamus at different levels of vigilance. *Cereb. Cortex* **7**, 583–604 (1997).
21. Destexhe, A., Contreras, D. & Steriade, M. Mechanisms underlying the synchronizing action of corticothalamic feedback through inhibition of thalamic relay cells. *J. Neurophysiol.* **79**, 999–1016. (1998).
22. Amzica, F. & Steriade, M. Electrophysiological correlates of sleep delta waves. *Electroencephalogr. Clin. Neurophysiol.* **107**, 69–83 (1998).
23. Alkire, M. T., Haier, R. J. & Fallon, J. H. Toward a unified theory of narcosis: brain imaging evidence for a thalamocortical switch as the neurophysiologic basis of anesthetic- induced unconsciousness. *Conscious. Cogn.* **9**, 370–86. (2000).
24. Thomschewski, A., Hincapié, A.-S. & Frauscher, B. Localization of the Epileptogenic Zone Using High

Frequency Oscillations. *Front. Neurol.* **10**, 94 (2019).
25. Srejic, L. R., Valiante, T. A., Aarts, M. M. & Hutchison, W. D. High-frequency cortical activity associated with postischemic epileptiform discharges in an in vivo rat focal stroke model. *J. Neurosurg.* **118**, 1098–1106 (2013).
26. Bragin, A. *et al.* Interictal high-frequency oscillations (80-500 Hz) in the human epileptic brain: entorhinal cortex. *Ann. Neurol.* **52**, 407–415 (2002).
27. Lizier, J. T. JIDT: An Information-Theoretic Toolkit for Studying the Dynamics of Complex Systems. *Frontiers in Robotics and AI* **1**, 11 (2014).
28. Yang, C. C., Shaw, F. Z., Lai, C. J., Lai, C. W. & Kuo, T. B. Relationship between electroencephalogram slow-wave magnitude and heart rate variability during sleep in rats. *Neurosci. Lett.* **336**, 21–24 (2003).
29. Kuo, T. B. J., Lai, C. T., Chen, C. Y., Yang, Y. C. & Yang, C. C. H. The high-frequency component of heart rate variability during extended wakefulness is closely associated with the depth of the ensuing sleep in C57BL6 mice. *Neuroscience* **330**, 257–266 (2016).